\documentclass[useAMS,usenatbib]{mn2e}
\usepackage{graphicx}
\usepackage{times}

\newcommand{\etal}{{et al.~}}
\newcommand{\beq}{\begin{equation}}
\newcommand{\eeq}{\end{equation}}
\newcommand{\Mo}{{\rm M_\odot}}
\newcommand{\kms}{\ {\rm km}\,{\rm s}^{-1}}
\def\LCDM{$\Lambda$CDM}
\def\kpc{\ {\rm kpc}}
\def\Mpc{\ {\rm Mpc}}
\def\lsim {\lower .1ex\hbox{\rlap{\raise .6ex\hbox{\hskip .3ex
        {\ifmmode{\scriptscriptstyle <}\else
                {$\scriptscriptstyle <$}\fi}}}
        \kern -.4ex{\ifmmode{\scriptscriptstyle \sim}\else
                {$\scriptscriptstyle\sim$}\fi}}}

\title[The origin and tidal evolution of cuspy triaxial haloes]
{The origin and tidal evolution of cuspy triaxial haloes}

\author[B. Moore \etal] {Ben Moore
\thanks{E-mail: moore@physik.unizh.ch (BM);
stelios@physik.unizh.ch (SK); diemand@physik.unizh.ch (JD), 
stadel@physik.unizh.ch (JS)},
Stelios Kazantzidis, J\"urg Diemand and Joachim Stadel  \\
Institute for Theoretical Physics, University of Z\"urich, CH-8057 Z\"urich,
Switzerland}

\begin{document}

\maketitle
\begin{abstract}
We present a technique for constructing equilibrium triaxial $N$-body haloes
with nearly arbitrary density profiles, axial ratios and spin parameters. The method is based
on the way in which structures form in hierarchical cosmological simulations, where prolate
and oblate haloes form via mergers with low and high angular momentum, respectively.
We show that major mergers between equilibrium spherical cuspy
haloes produce similarly cuspy triaxial remnants and higher angular-momentum 
mergers produce systems with lower concentrations. Triaxial haloes orbiting within 
deeper potentials become more spherical and their velocity dispersion tensors more isotropic.
The rate of mass loss depends sensitively on the halo shape: a prolate 
halo can lose mass at a rate several times higher than an isotropic spherical 
halo with the same density profile. Subhaloes within cosmological simulations are 
significantly rounder than field haloes with axial ratios that are $\sim30\%$ larger.

\end{abstract}


\section{INTRODUCTION}

A generic prediction of the currently favoured cold dark matter (CDM)
cosmological model of hierarchical structure formation is that dark matter 
(DM) haloes of galaxies and clusters are flattened triaxial systems 
\citep[e.g.,][]{barnes_efstathiou87,frenk_etal88,dubinski_carlberg91,warren_etal92,
cole_lacey96,thomas_etal98,jing_suto02}.
Observationally, inferring the intrinsic shape of DM haloes of galaxies and clusters
is a difficult task, but nonetheless several potentially powerful probes exist 
that may allow us to to distinguish between spherical and flattened DM haloes.
These include the dynamical modelling of collisionless tracers such as 
tidal streams orbiting the Milky Way 
\citep{johnston_etal99,ibata_etal01,mayer_etal02,majewski_etal04}, 
the distribution and kinematics of gas in spiral galaxies 
\citep{kuijken_tremaine94,franx_etal94,schoenmakers_etal97,merrifield_02}, 
and polar ring galaxies \citep{schweizer_etal83,sackett_sparke90,
sparke02,iodice_etal03} which provide shape constraints perpendicular to 
the disc plane, gravitational lensing applications 
\citep{kochanek95,bartelmann_etal95,koopmans_etal98,oguri_etal03} 
and the flattening of the extended X-ray isophotes in elliptical galaxies 
\citep{buote_canizares94,buote_canizares96,buote_canizares98,buote_etal02}.

Numerical experiments of isolated equilibrium models are very useful
for studying the dynamical evolution of gravitating systems in a 
controlled way. Several techniques exist for constructing $N$-body realizations 
of spherical haloes and multi-component galaxies
\citep[][hereafter KMM]{hernquist93,bockelmann_etal03,kazantzidis_etal04a}, 
but it is far more complicated to build triaxial equilibria.
Even though the Jeans theorem guarantees that the distribution function 
depends only on the isolating integrals of motion, 
explicit expressions for the latter other than the energy  
per unit mass, $E$, are rarely known in the case of triaxial potentials.
As a result, the triaxial models that have been constructed so far
have been either limited to a few special analytical cases (e.g., 
St\"ackel potentials or rotating $f(E_{J})$ models with $E_{J}$ being the 
Jacobi constant) or entirely based on numerical techniques 
\citep{schwarzschild79,schwarzschild93,merritt_fridman96,poon_merritt01,terzic03}.

Recently, \citet{bockelmann_etal01} used a technique of adiabatically applying a drag 
to the velocities of the particles along each principal axis in order to create
cuspy triaxial systems starting with a spherical Hernquist \citep{hernquist90} model. 
The original prescription of \citet{hernquist93} was generalized
by \citet{boily_etal01} for accomodating composite, axisymmetric models of galaxies
and extended by \citet{tinker_ryden02} for investigating the 
effect of rotating, triaxial halos on disk galaxies. 
These techniques are useful and have been used to understand the response of
systems to live triaxial potentials. However, they are somewhat 
restricted to modest values of the flattening and it is difficult to 
incorporate a significant amount of rotational flattening. 

Triaxiality can arise in a number of ways. The most common way of 
constructing numerical models of triaxial galaxies is via the merger of other objects. 
Examples include binary mergers of spherical haloes 
\citep[e.g.,][]{white78,fulton_barnes01}
and disk galaxies \citep[e.g.,][]{gerhard_81,barnes92,barnes_hernquist96,naab_burkert03,
kazantzidis_etal04b}, as well as multiple mergers of systems 
\citep[e.g.,][]{weil_hernquist96,dubinski98}.
The structure of the final remnant of two component models depends sensitively on
both the orbital geometry \citep{naab_burkert03} as well as on the inclination and 
internal properties of the disks \citep{kazantzidis_etal04b}.

A general technique for constructing cuspy axisymmetric and triaxial $N$-body 
systems would be advantageous for many purposes including studying the 
dynamical friction in flattened and 
rotating systems, the mass loss and tidal stripping from triaxial 
substructure haloes, the effects of baryonic accretion and disc
formation on triaxial halo shape and anisotropy, the effects of triaxial shape on the
properties and stability of discs, the interaction between central black holes and 
cusps and the inflow of gas in triaxial systems.

In this paper we explore the generation of triaxial structures that is based on the 
way that haloes in hierarchical models of structure formation obtain their shapes.
Inspection of a cosmological simulation 
reveals that triaxiality arises via mergers that take place with various amounts of
angular momentum which is generated from the large-scale
tidal field. Mergers between two haloes that occur with little 
angular momentum (radial mergers) produce prolate systems, whilst high
angular momentum mergers produce oblate systems.  
Most CDM haloes form via a sequence of mergers with varying amounts of 
angular momentum such that haloes with arbitrary triaxiality may be formed and 
the triaxiality will vary with radius \citep{moore_etal01,vitvitska_etal02}.

The outline of this paper is as follows.
In Section~2, we describe our technique for constructing 
equilibrium cuspy $N$-body haloes with various degrees of flattening. 
There we discuss in detail the numerical experiments we performed and 
present our results for the internal structure of the resulting models.
In Section~3, we investigate the tidal evolution of triaxial 
substructure haloes within a static host potential and 
study the shape of subhaloes within a cosmological CDM simulation. 
Finally, we summarise our main results in Section~4.

\section{CONSTRUCTING CUSPY TRIAXIAL DARK MATTER HALOES}

We begin with a cuspy spherical DM halo which follows the 
\citet[hereafter NFW]{navarro_etal96} density profile
\beq
\rho(r)=\frac{\rho_{\rm s}} {(r/r_{\rm s}) (1+r/r_{\rm s})^2} 
\qquad\hbox{($r \leq r_{\rm vir}$)}.
\label{NFW_profile}
\eeq 
Here the characteristic inner density $\rho_{\rm s}$ and scale radius 
$r_{\rm s}$, are sensitive to the epoch of halo formation and tightly
correlated with the halo virial parameters, via the concentration, $c$, and 
the virial overdensity $\Delta_{\rm vir}$.
Since the NFW density profile corresponds to a
cumulative mass distribution that diverges as $r\rightarrow\infty$,
we impliment an exponential cut-off for $r > r_{\rm rvir}$ which sets
in at the virial radius and turns off the profile
on a scale $r_{\rm decay}$ which is a free parameter and controls the
sharpness of the transition
\beq
\rho(r)=\frac{\rho_{\rm s}} {c (1+c)^2} 
\left(\frac{r}{r_{\rm vir}}\right)^{\epsilon}
\exp\left[-\frac{r-r_{\rm vir}}{r_{\rm decay}}\right]  
(r>r_{\rm vir}) \ ,
\label{exp_cutoff}
\eeq
where $c\equiv r_{\rm vir}/r_{\rm s}$ is the concentration parameter.
Finally, in order to ensure a smooth transition between (\ref{NFW_profile})
and (\ref{exp_cutoff}) at $r_{\rm vir}$, we require the logarithmic slope 
there to be continuous. This implies 
\beq 
\epsilon=-\frac{1+3c}{1+c} +
\frac{r_{\rm vir}}{r_{\rm decay}}.
\label{eps}
\eeq 
\begin{figure*}
\begin{center}
\resizebox{8cm}{!}{\includegraphics{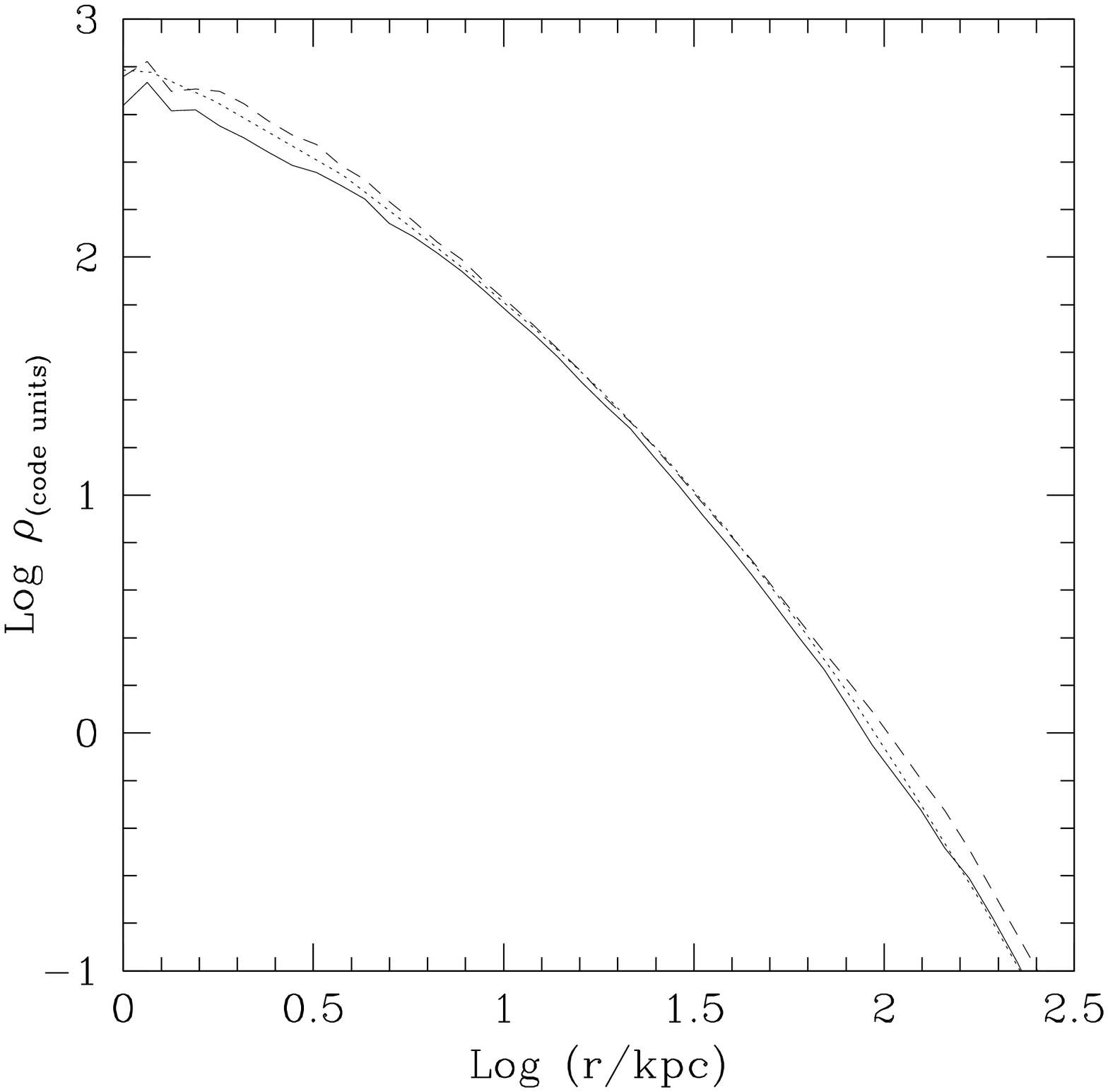}}
\resizebox{7.8cm}{!}{\includegraphics{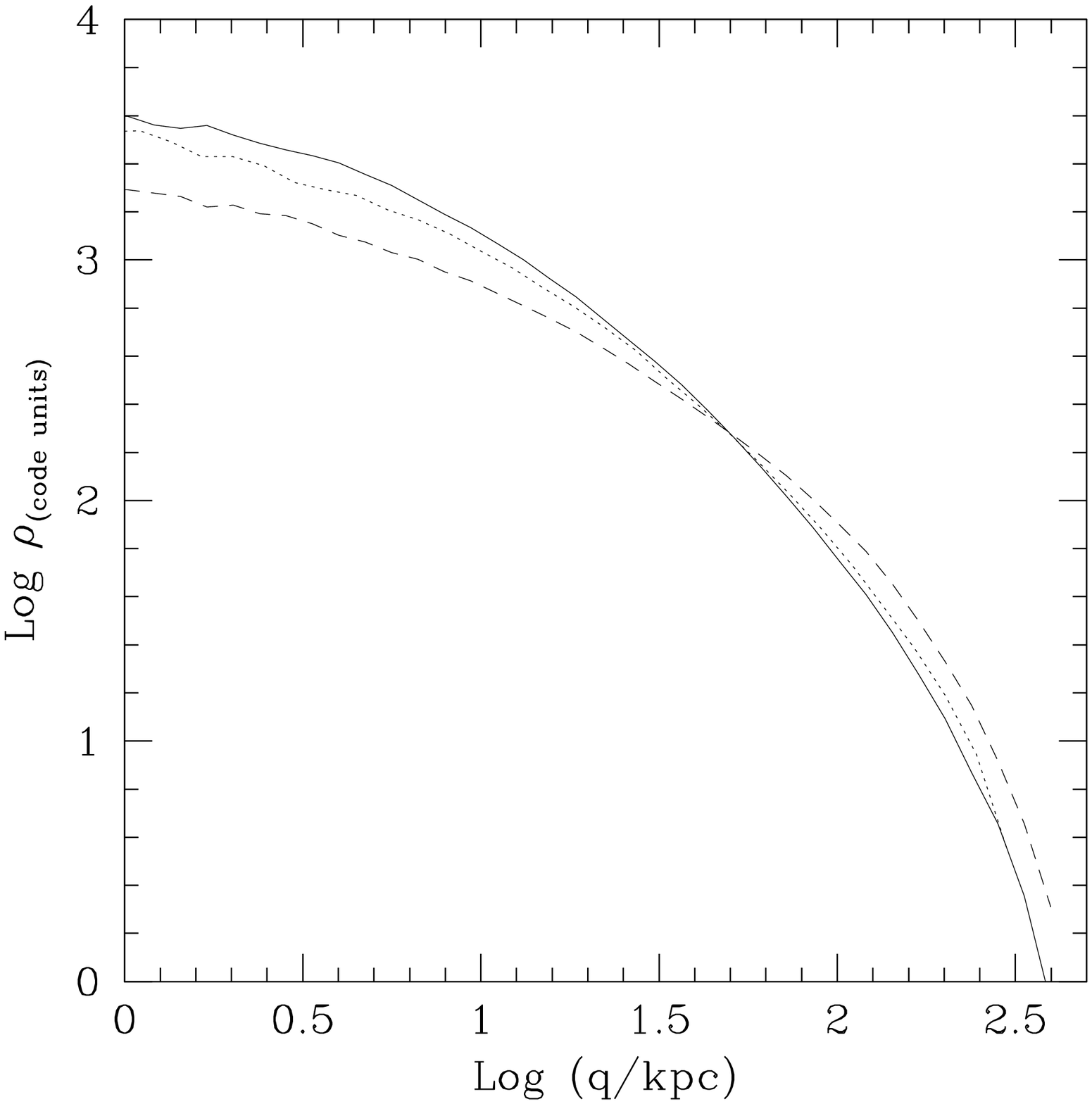}}\\
\caption{Density profiles of the initial model (dotted curve),
the $a$:$b$:$c$ = $2$:$1$:$1$ prolate model (solid curve) and the 
$a$:$b$:$c$ = $2$:$2$:$1$ oblate model (dashed curve). 
The density was calculated using spherically averaged 
(left panel) and ellipsoidally averaged (right panel) bins such that the density is 
projected onto the symmetry axis.
\label{fig1}}
\end{center}
\end{figure*}
\begin{figure*}
\begin{center}
\resizebox{8cm}{!}{\includegraphics{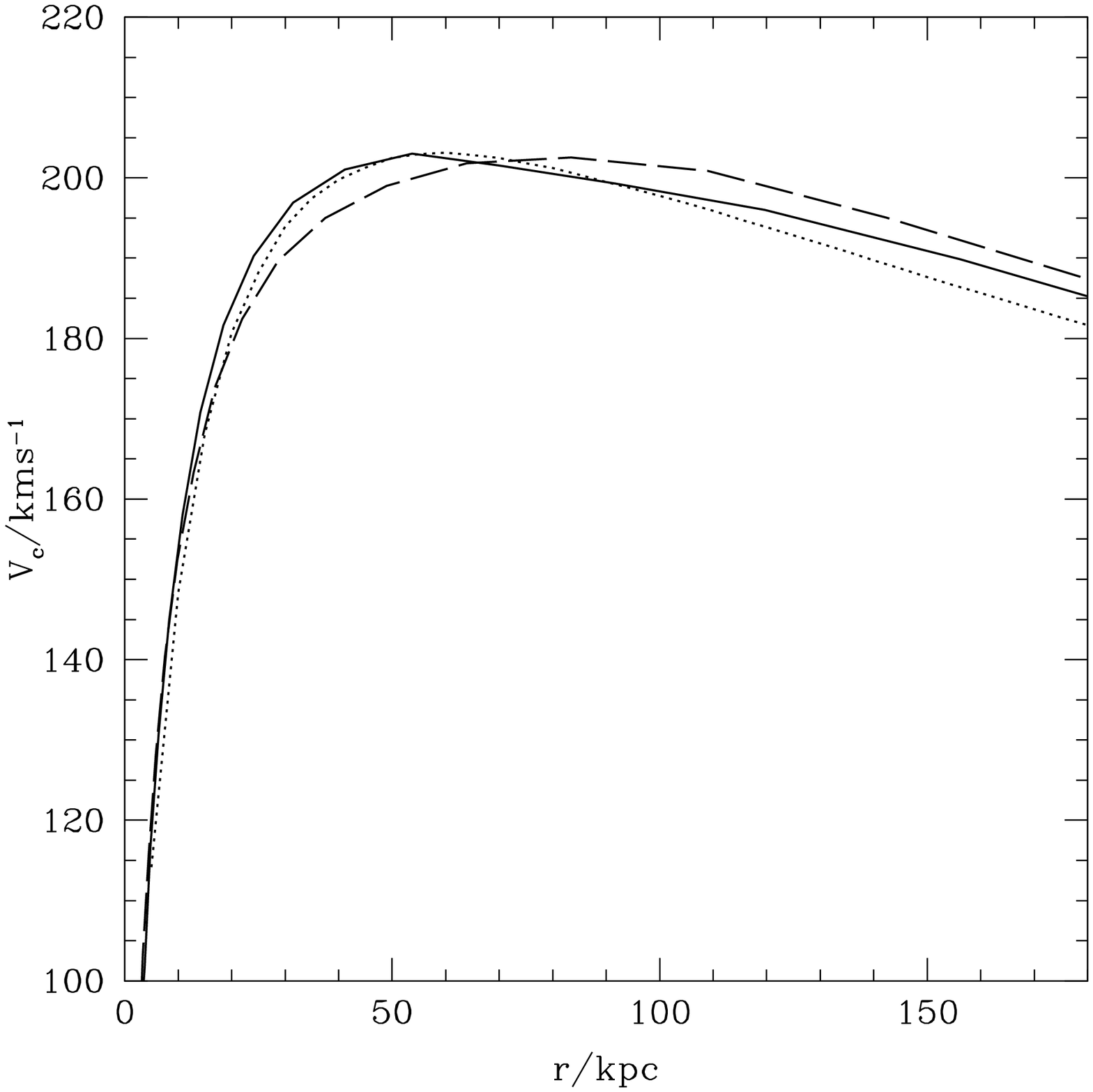}}
\resizebox{8cm}{!}{\includegraphics{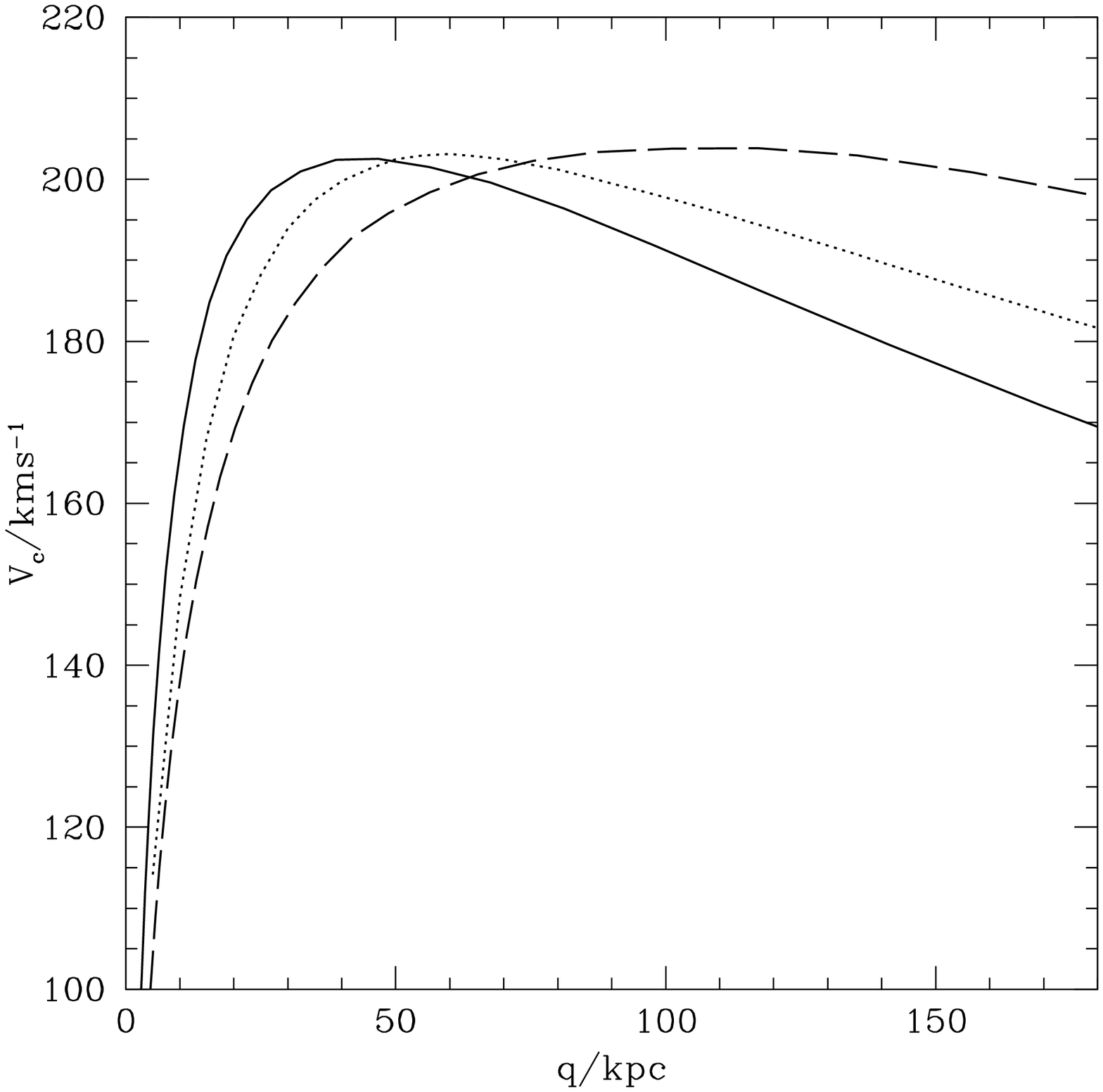}}\\
\caption{Circular velocity curves scaled to the same peak circular velocity for the initial 
model (dotted curve), prolate $a$:$b$:$c$ = $2$:$1$:$1$ halo 
(solid curve) and oblate $a$:$b$:$c$ = $2$:$2$:$1$ halo (dashed curve).
The mass distribution was calculated using spherically averaged 
(left panel) and ellipsoidally averaged (right panel) bins.
In each case the circular velocities are scaled to the same peak velocity assuming
$V_{\rm c}\propto r$.
\label{fig2}}
\end{center}
\end{figure*}

Once the density profile $\rho(r)$ of our initial model has been specified,
we construct an $N$-body realization using the technique discussed in
detail in KMM. To summarize, the particle positions and velocities
are initialized by sampling the 
exact phase-space distribution function using Eddington's inversion formula 
\citep{eddington16} for the given density profile $\rho(r)$
and assuming spherical symmetry.
Note that in this paper we will consider only initial models with isotropic 
velocity dispersion tensors and therefore their distribution function
will only depend on the binding energy $E$, $f=f(E)$.
\begin{figure*}
\includegraphics[width=16cm]{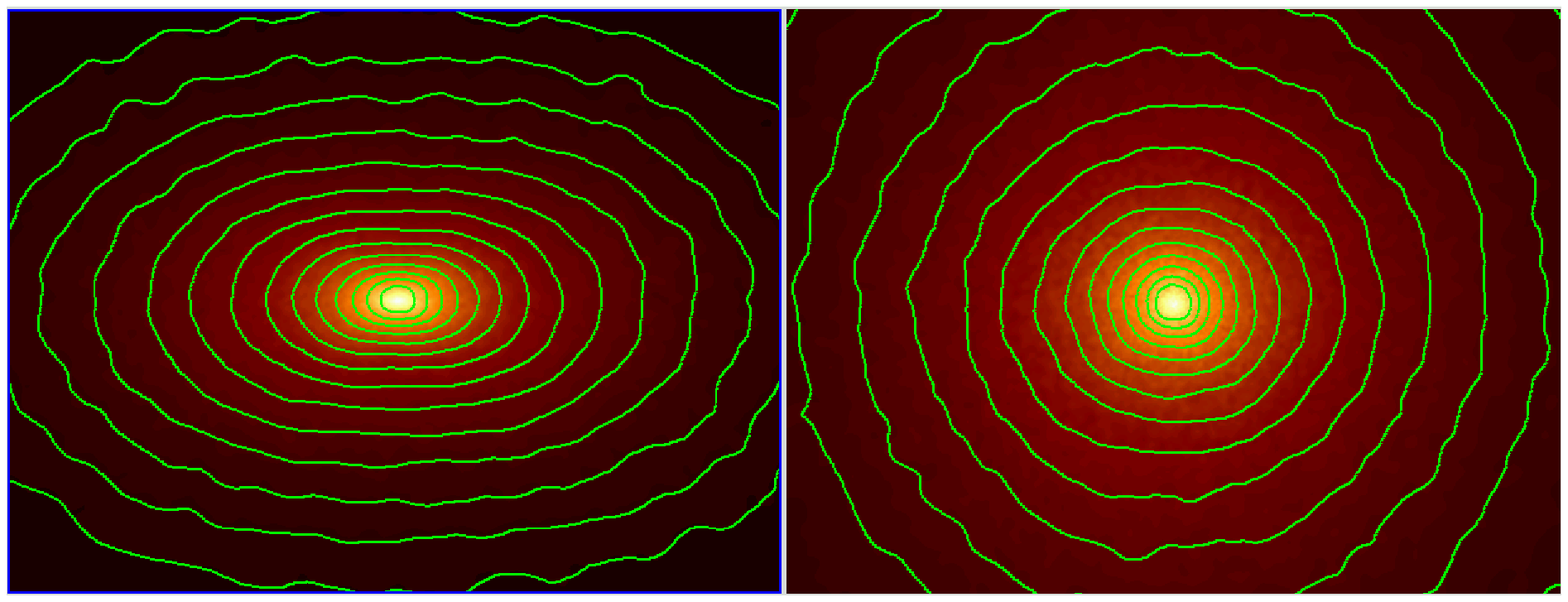}
\caption{Logarithmically spaced surface density contours of the oblate model 
viewed in the plane of the long axis (left) and short axis (right).
\label{fig3}}
\end{figure*}
For the following experiments we begin with an initial galaxy-sized halo 
with virial mass $M_{\rm vir}=7 \times10^{11} \,h^{-1} \Mo$ 
and concentration of $c=10$ in accordance with theoretical 
predictions for objects at this mass scale and for the adopted cosmology 
\citep[e.g.,][]{bullock_etal01}. 
Note that for all simulations presented in this paper including the cosmological 
ones described in the next section, we consider as our framework the 
concordance flat $\Lambda$CDM model with present-day matter and 
vacuum densities $\Omega_{\rm m}=0.3$
and $\Omega_{\Lambda}=0.7$, respectively, dimensionless Hubble
constant $h=0.7$, present-day fluctuation amplitude $\sigma_{8}=0.9$ and
index of the primordial power spectrum $n=1.0$.
For this choice of concentration, the virial and scale radius of the initial
halo model are $r_{\rm vir}=181 \, h^{-1} \kpc$ and $r_{\rm s}=18.1 \, h^{-1} \kpc$,
respectively. The initial model is constructed 
with $N=5\times10^{5}$ particles and we use 
a gravitational softening of $\epsilon=0.3 \, h^{-1} \kpc$.

All the simulations presented in this paper were carried out  
using PKDGRAV, a multi-stepping, parallel
$N$-body code \citep{stadel01} which
uses a spline kernel softening and multi-stepping based on the
local acceleration of particles.
The time integration was performed with 
high enough accuracy such that the total energy was conserved 
to better than 0.1\% in all cases.
We have also explicitly checked that our results are not compromised 
by choices of force softening, time-stepping or opening angle criteria 
in the treecode.

We perform a first experiment in which we merge two identical spherical haloes with 
zero angular momentum corresponding to a head-on collision with an impact
parameter $b=0$. The initial relative separation of the two haloes is 
$r_{\rm sep}=210h^{-1}\kpc$ and are they infalling at a relative speed of 
$v_{\rm rel}=300\kms$. Relative velocities 
significantly higher than this value lead to unbound orbits, so we 
chose this value so as to maximize the amount of flattening from a single merger.
Note that the initial separation corresponds to haloes which are overlapping. 
This does not influence the final result as nearly identical
remnants were produced with non-overlapping initial models.
The two haloes merge within a time-scale of order the crossing
time $t_{\rm cross}=\sqrt{r_{\rm vir}^{3} / G M_{\rm vir}}$ 
which is approximately equal to $2$ Gyr for our initial models.
In all simulations the merger remnants were allowed to settle into equilibrium for several 
crossing times after the merger was complete which was established by monitoring 
the fluctuations of the density profile of the final system. In addition, the virial ratio 
$2T/W$ varied between 0.1 and 0.3\% from $1.0$ for the same time-scales confirming 
that our triaxial remnants are indeed in equilibrium. 
In all cases the centre of the remnant is identified using the most bound particle, 
which agrees very well with the centre of mass recursively calculated using smaller 
spherical regions. 

The radial merger produces a prolate axisymmetric system with
axial ratios $a$:$b$:$c$ = $2$:$1$:$1$ where $a$, $b$ and $c$ are the long, 
intermediate and short axis, respectively. 
For each remnant, we calculate principle axis ratios $s = b/a$ and 
$q =c/a$ ($a>b>c$), from the eigenvalues of a 
modified dimensionless inertia tensor \citep[e.g.,][]{dubinski_carlberg91}: 
$I_{ij} = \sum x_{i}x_{j}/q^{2}$, where $q^2=x^2 + (y/s)^2 + (z/q)^2$
is the ellipsoidal coordinate. We use an iterative algorithm starting with a spherical
configuration ($a=b=c$), and use the results of the previous
iteration to define the principle axes of the next iteration. The procedure is iterated 
until the values of both axial ratios $b/a$ and $c/a$ have a percentage change of less
than $10^{-3}$. The dimensionless spin parameter of this remnant, which 
is defined by
\beq
\lambda=\frac{J |E|^{\frac{1}{2}}}{G M^{\frac{5}{2}}}
\eeq
where $J$, $E$ and $M$ are the total angular momentum, binding energy and mass
respectively, corresponds to $\lambda=0$ as expected in a radial merger.

\begin{figure}
\includegraphics[width=8cm]{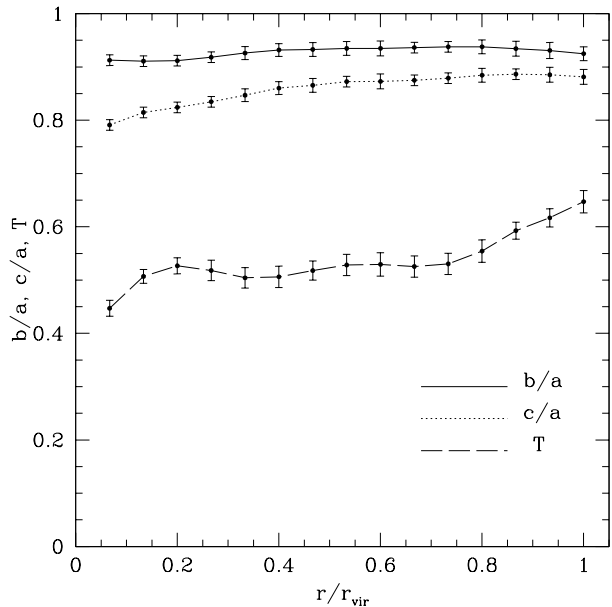}
\caption{Axial ratios and triaxiality parameter of the minor merger remant. 
The solid, dotted and dashed lines, respectively, show results for $b/a$, $c/a$ and 
$T$. The profile of the triaxiality parameter shows that this remnant is strongly triaxial
with $T\sim 0.5$. The error bars represent the {\sl rms} variation in the measurements from 
$20$ time intervals spanning $10$ half-mass dynamical times of the remnant 
at the end of the simulation. The errors in the axial ratios and triaxiality parameter are 
small demonstrating that the remnant is in equilibrium.
We define the virial radius of the remnant, $r_{\rm vir}$, to be the 
distance from the centre where the mean enclosed density is $\sim 337$ times
the mean universal density.
\label{fig4}}
\end{figure}

In a second experiment we start with the two spherical haloes
separated again by $r_{\rm sep}=210h^{-1}\kpc$, but we give one halo
a transverse velocity equal to the circular velocity of the combined model
at that distance, $v_{\rm rel}=240\kms$.
The circular orbit merger produces an oblate halo with axial ratios 
$a$:$b$:$c$ = $2$:$2$:$1$.
As we demonstrate later (see Figure~\ref{fig4}), haloes with any amount of triaxiality 
between these values could be created by using unequal-mass mergers or mergers 
with less or more angular momentum. The spin parameter of the final system is 
$\lambda=0.1$. As expected, with these experiments we have been able to reproduce most 
of the range in $\lambda$ seen in cosmological $N$-body simulations 
\citep[e.g.,][]{warren_etal92,lemson_kauffmann99}. Although the formation
histories of halos with a hierarchical universe is significantly more complex
than these binary mergers, we might expect that oblate halos have a higher
spin parameter and more angular momentum than prolate halos.
Figure~\ref{fig1} shows the density profiles of the initial spherical model and that of 
the prolate and oblate haloes. The panel on the left shows results 
using spherically-averaged bins. The density profiles on the right panel 
were calculated as a function of the ellipsoidal coordinate $q$, where 
$q^2=x^2 + (y/b)^2 + (z/c)^2$. In agreement with other studies 
\citep[e.g.,][]{white78,villumsen83,barnes99,fulton_barnes01,boylan_ma04},
remnant central density slopes remain unchanged.
In Figure~\ref{fig2} we show the circular velocity curves of the initial model and the
final prolate and oblate haloes using again spherically averaged 
(left panel) and ellipsoidally averaged (right panel) bins. These have all been scaled by 
the same factor in radius and velocity such that all the curves have the same peak 
circular velocity. The prolate halo has a nearly identical concentration to the initial 
spherical model, but the oblate halo is nearly a factor of two less concentrated.

In order to create haloes with higher amounts of flattening it is necessary to
merge together the products of the first set of simulations. Merging two prolate
haloes along the long axis again at a velocity such that the system is just bound 
produces a $a$:$b$:$c$ = $3$:$1$:$1$ remnant, while merging two oblate haloes in the 
plane of the long axis, but counter-rotating, 
results in a further flattened system with axial ratios $a$:$b$:$c$ = $3$:$3$:$1$
and no net angular momentum, but with particles streaming in opposite directions.
Figure~\ref{fig3} shows two contoured images of the oblate halo viewed
projected along the short and long axis, respectively. 

Finally, we performed a minor merger between two spherical halos with mass 
ratio equal to $4:1$. The most massive halo was identical to the ones used 
in the previous experiments while the concentration of the smaller one was 
scaled according to \citet{bullock_etal01} for the adopted {\LCDM} cosmology.
The orbital angular momentum of the merger was typical of cosmological 
encounters \citep{khochfar_burkert03}.
The number of particles and gravitational softening used for the less massive halo 
was such that both halos were resolved with the same mass and force resolution.
This is useful in the intepretation of the results regarding the structure of the 
remnant. \citet{kazantzidis_etal04b} examined the evolution of disk galaxy merger
remants and found that their shapes evolve in the outer regions for several 
crossing times. In Figure~\ref{fig4}, we demonstrate the stability of this remnant
by running the simulation for a further $10$ half-mass dynamical times and 
calculated profiles of axial ratios and triaxiality parameter 
[$T\equiv (a^{2}-b^{2})/(a^{2}-c^{2})$] at $20$ intermediate steps.
The error bars in Figure~\ref{fig4} show the  {\sl rms} variation in the profiles 
at each radius and demonstrate that the remnant is in equilibrium.
\begin{figure}
\includegraphics[width=8cm]{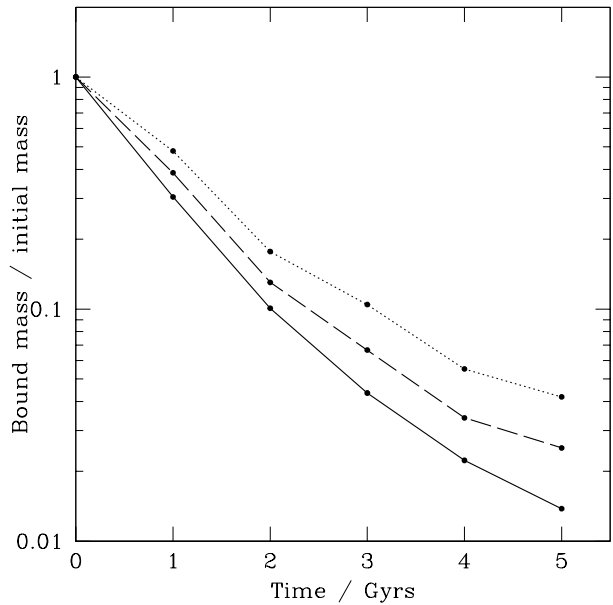}
\caption{The rate of tidal mass loss over a period of $5$ Gyr 
($\sim 2$ orbital periods) for the initial spherical model 
(dotted line), the $a$:$b$:$c$ = $3$:$1$:$1$
prolate model (solid curve) and the $a$:$b$:$c$ = $3$:$3$:$1$
oblate model (dashed curve). The prolate halo experiences much more efficient 
tidal stripping than both the spherical and oblate halo 
on the same external tidal field and orbit which is probably caused
by the orbital distribution of particles supporting its shape.
\label{fig5}}
\end{figure}
\begin{figure}
\includegraphics[width=8cm]{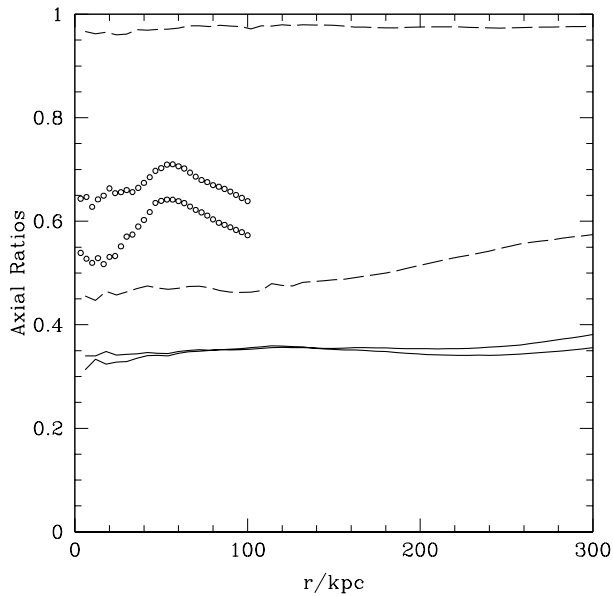}
\caption{The axial ratios as a function of radius for the prolate (solid lines)
and oblate (dashed lines) $3$:$1$ haloes.
The open circles show the axial ratios of the same prolate model after 
orbiting for $5$ Gyr within a deeper potential.
\label{fig6}}
\end{figure}
\begin{figure}
\includegraphics[width=8cm]{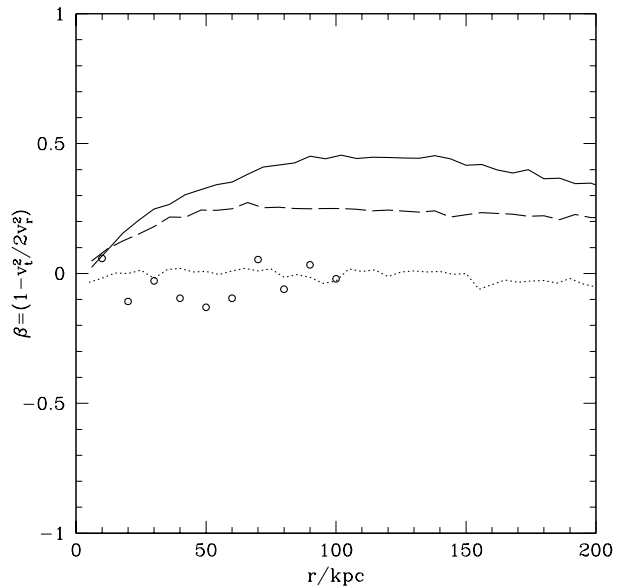}
\caption{The anisotropy parameter $\beta$ as a function of radius 
for the prolate halo model (solid line), oblate halo (dashed line) and initial 
conditions (isotropic spherical halo, dotted line). 
The open symbols show the anisotropy parameter for the prolate halo after orbiting
for $5$ Gyr within a deeper potential.
\label{fig7}}
\end{figure}

\section{TIDAL EVOLUTION OF TRIAXIAL SUBHALOES} 

Numerical investigations of the tidal stripping and mass loss of substructure haloes 
in CDM models have previously been restricted to spherical systems with isotropic 
velocity dispersion tensors \citep[e.g.,][]{taffoni_etal03,hayashi_etal03}. 
The results of these simulations are also incorporated into semi-analytic models 
which attempt to follow the tidal evolution of orbiting subhaloes. 
However, the shapes and velocity dispersion tensors of haloes in cosmological 
simulations have been studied by many authors 
\citep[e.g.,][]{warren_etal92,cole_lacey96,thomas_etal98,colin_etal00,bullock02}
and they exhibit a significant departure from spherical symmetry and isotropy.
It is interesting to investigate if the response of a spherical subhalo to an imposed tidal
field is the same as that of a triaxial subhalo.

KMM compared the resilience to tidal forces
of two identical self-consistent NFW satellites on the same external 
tidal field and orbit, but constructed with different velocity dispersion tensors.
Identifying the bound mass as function of time, these authors demonstrated 
that the satellite with a radially anisotropic velocity dispersion  
\citep{osipkov79,merritt85} experiences much more 
efficient mass loss than its isotropic counterpart. This is due to the fact that 
particles on more radial orbits spend on average more time at larger radii and 
are therefore more easily stripped by the external field. 
\begin{figure*}
\includegraphics[width=16cm]{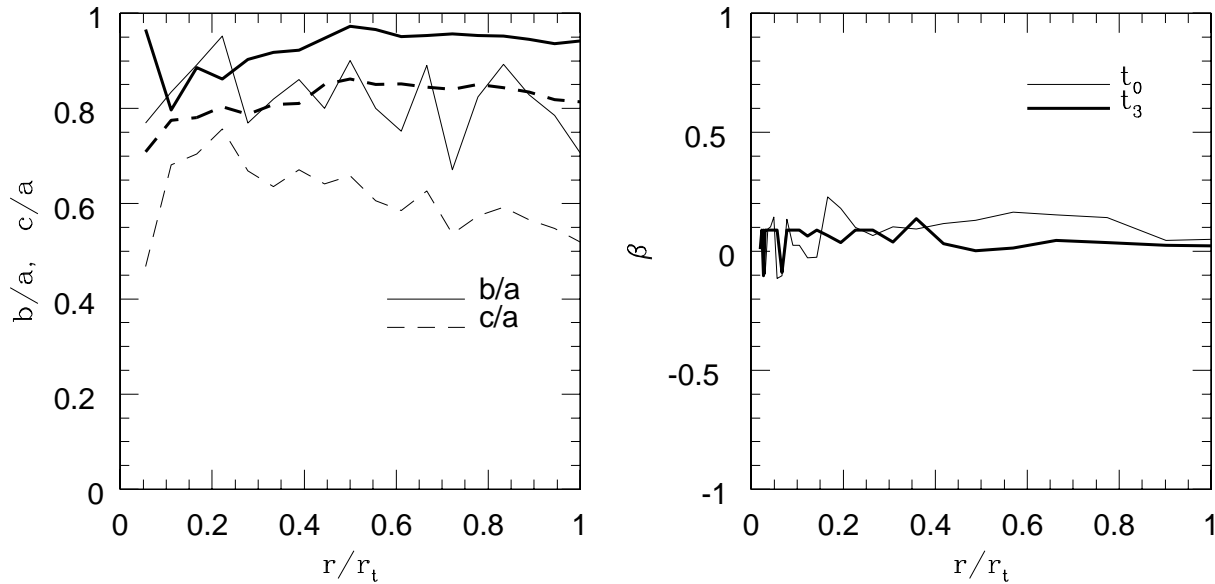}
\caption{Structural and kinematic properties of one subhalo 
in our highest-resolution cosmological run (R12) resolved  
with more than $2 \times10^5$ particles.
Left: The axial ratios $b/a$ and $c/a$ as a function of radius from the 
centre of the subhalo. Right: The anisotropy parameter $\beta$ as a function 
of radius. In both cases, the thin lines correspond to measurements
of the subhalo's properties just before entering the primary halo, whereas
the thick lines correspond to its third pericentric passage.
The subhalo becomes more spherical and isotropic as a result of the tides.
Note that the radius of the subhalo is given in units of $r_{\rm t}$, where 
$r_{\rm t}$ is its tidal radius after the third pericentre passage.
\label{fig8}}
\end{figure*}
Here we expand on the previous result and show that similar differences are obtained 
when one considers the tidal evolution of spherical and triaxial haloes as they orbit 
within a deeper potential. We compare results of the $3$:$1$ prolate and oblate haloes and 
the initial isotropic spherical halo used to create these models. The density profile
of the spherical model is scaled by a factor of four in mass to preserve the same 
spherically averaged density profile as the prolate halo. Each halo is then placed
on an identical orbit within a static cored isothermal potential such that 
the ratio of host to satellite halo circular velocities is 
$v_{\rm host}/v_{\rm sat}$ = $10$:$1$ and the apocentric
to pericentric radii is $r_{\rm apo}/r_{\rm per}$ = $4$:$1$. 
The alignment of the axes of the triaxial satellite were varied with respect to the 
external deeper potential and the satellites always begin at the apocentre which is set 
equal to the virial radius. This allows the satellite-host system to be rescaled to any 
galactic or cluster system.  
The imposed potential is reasonably soft since we use a core
radius equal to half the pericentric radius. Significantly higher rates of mass loss
may be expected in a more cuspy potential, however here we aim to explore the
slow tidal loss of mass rather than mass loss due to strong tidal shocks. In a 
forthcoming paper we will give an analytical model for the rate of mass loss 
as a function of host potential, orbit and subhalo structure and shape 
(Kazantzidis \etal, in preparation).

In Figure~\ref{fig5} we perform a comparison of the tidal evolution of each 
of these three models by identifying the bound mass as a function of time
using the publicly available group 
finder SKID\footnote {http://www-hpcc.astro.washington.edu/tools/skid.html} \citep{stadel01}.
The prolate halo experiences mass loss at a much higher rate
than both the spherical and oblate ones. We also found that this result
is not significantly sensitive to the alignment between the major or minor axis of the 
satellite and the orbital plane with the maximum differences in the mass loss being of
the order of 25\%. In Figures~\ref{fig6} and~\ref{fig7} we show the change in the
shape and in the anisotropy parameter $\beta$ of our models.
We suspect that the prolate halo becomes more spherical due to the tidal removal of particles 
on radial orbits that are supporting the shape of the system. Future studies will analyse
the orbital evolution of particles within the satellite to study these effects in detail.
In addition, after $5$ Gyr of orbital evolution the velocity dispersion tensor of the prolate 
halo has significantly changed and now is nearly isotropic.

It is interesting to study the evolution of shape and anisotropy for 
haloes  in a CDM simulation that enter into a deeper potential. 
This is important when we want to model the orbits of 
stars in satellite galaxies or the effect of subhaloes on lense models.
Here we analyse subhaloes from a high resolution 
$\Lambda$CDM cluster simulation \citep{diemand_etal04}.
The initial conditions are generated with
the GRAFIC2 package \citep{bertschinger01}.
We start with a $300^3$ particle cubic grid with
a comoving cube size of $300 \Mpc$ (particle mass 
$m_{\rm p} = 2.6\times 10^{10}h^{-1} \Mo$).
We trace back and refine a cluster region
by a factor of $12$ in length and $1728$ in mass, 
so that the mass resolution is $m_{\rm p} = 1.5 \times 10^7 h^{-1}\Mo$.
The softening length is comoving
from the start of the simulation ($z \simeq 40$) to $z = 9$.
From $z = 9$ until present, we use a physical softening length of 
$1.05 \times 10^{-3} r_{\rm vir,parent}$. At $z = 0$ the refined cluster contains 
$1.4\times 10^7$ particles (R12) and we select a relatively massive subhalo
which is resolved with more than $2 \times10^5$ particles.

Figure~\ref{fig8} shows the axial ratios $b/a$ and $c/a$ (left panel)
and the anisotropy parameter $\beta$ (right panel) as a function 
of radius for this object.
The thin lines correspond to the subhalo just before entering its host, whereas the 
thick lines correspond to the third pericentre passage of the subhalo.
Note that the radius of the subhalo is given in units of $r_{\rm t}$, where 
$r_{\rm t}$ is its tidal radius after the third pericentre passage.
Figure~\ref{fig8} confirms that subhaloes become more spherical
and their velocity distribution more isotropic owing to tidal effects.
We then compare the mean axial ratios of field haloes versus subhaloes
in the same simulation. On average we find that subhaloes have axial
ratios that are 30\% larger (more spherical) than field haloes.

\section{CONCLUSIONS}

We have presented a method for constructing cuspy axisymmetric 
and triaxial $N$-body haloes based on merging isotropic equilibrium spherical 
haloes with varying amounts of angular momentum.
In particular, we found that radial mergers produce prolate systems, while mergers 
on circular orbits produce oblate systems. 
This technique has the benefit that it is based on the 
way in which haloes obtain their triaxiality in cosmological simulations, 
therefore the anisotropy distributions and angular momentum of haloes are 
well motivated. 

Mergers between similar equilibrium spherical haloes at high resolution show 
that the resulting halo has the same density profile, 
independent of the angular momentum of the merger.
This has two implications: (i) the density profile of the triaxial halo can be
set by the choice of the density profiles of the progenitor haloes and (ii) similar
mass mergers can not dramatically re-arrange the central 
density structure of DM haloes, albeit oblate haloes 
(high angular momentum merger remnants) have concentrations up to a 
factor of two lower than prolate haloes (low angular 
momentum merger remnants). This may explain the entire scatter in the distribution of halo 
concentrations from cosmological simulations 
\citep{bullock_etal01,eke_etal01,wechsler_etal02}.
It may also be the case that the least
concentrated haloes host the low surface brightness 
discs since these galaxies may form in high angular momentum
oblate haloes \citep[e.g.,][]{onell_etal97}. These galaxies generally require low
values of the concentration when fit to cuspy halo models 
\citep[e.g.,][]{mcgaugh_deblok98,van den Bosch_etal00,deblok_etal01,swaters_etal03}.

As an application we considered the tidal evolution of triaxial subhaloes orbiting 
within deeper potentials. Haloes with identical spherically averaged density profiles 
and on identical orbits evolve self-similarly with time. 
Spherical haloes with isotropic velocity dispersion tensors suffer 
significantly less mass loss than radially anisotropic prolate ones. This is 
likely due to the fact that particles on radial orbits are easily stripped which also 
results in subhaloes becoming more spherical.
In general, the subhaloes become more spherical
and their velocity distribution more isotropic owing to tidal effects.
This result has been confirmed by investigating the response to
tides of both isolated satellite haloes orbiting within a static host potential and 
substructure haloes in the time dependent cosmological tidal field. 
We find that subhaloes in a
cosmological simulation of a cluster are on average 30\% rounder than their
field counterparts. Galaxy subhaloes should be even closer to spherical since they 
spend longer being reshaped by the host potential. 

Our more realistic modelling of DM haloes is important for studies of the 
weak and strong lensing statistics attempting to distinguish between
competing cosmological models, the structural evolution and mass loss from 
substructure, the formation of tidal streams or the sizes of satellite haloes 
in galaxies and clusters. 
These results may also be important to incorporate within semi-analytic models
that attempt to model the distribution of satellites in DM haloes.
CDM haloes are generally radially anisotropic therefore they lose mass
and are disrupted more quickly than the isotropic systems that have been
generically considered in previous studies. This may explain why semi-analytic
models predict far more substructures than found in high resolution numerical
studies \citep[e.g.,][James Taylor private communication]{zentner_bullock03}.

\section*{Acknowledgments}

It is a pleasure to thank the anonymous referee for constructive comments
on the manuscript and Victor Debattista and Lucio Mayer for useful discussions.
BM thanks Priyamvada Natarajan for organising the Yale Cosmology Workshop
``The Shapes of Galaxies and Their Dark Matter Haloes'' (2001)
which motivated some of this work and also apologises for not writing 
up the conference proceedings. 
The numerical  simulations were carried out on the zBox 
(http://www-theorie.physik.unizh.ch/~stadel/). Initial conditions
for the cosmological simulations were generated at the 
Swiss Center for Scientific Computing (SCSC) at Manno.


\bsp

\label{lastpage}

\end{document}